\documentclass[12pt,preprint]{aastex}

\newcommand{\myemail}{zhouxu@bao.ac.cn}
\usepackage{url}
\usepackage{multirow}
\usepackage{graphics}
\usepackage{rotating}

\shorttitle{stellar populations and evolution of NGC 628}
\shortauthors{Zou et al.}

\begin{document}

\title{Stellar Population Properties and Evolution Analysis of NGC 628 with the Panchromatic Photometry}
\author{Hu Zou\altaffilmark{1,2}, Wei Zhang\altaffilmark{1}, Yanbin Yang\altaffilmark{1},
Xu Zhou\altaffilmark{1}, Zhaoji Jiang\altaffilmark{1}, Jun Ma\altaffilmark{1},
Zhenyu Wu\altaffilmark{1}, Jianghua Wu\altaffilmark{1}, Tianmeng Zhang\altaffilmark{1},
Zhou Fan\altaffilmark{1}}
\altaffiltext{1}{National Astronomical Observatories, Chinese Academy of
Sciences, Beijing 100012, China; \myemail}
\altaffiltext{2}{Graduate University of Chinese Academy of Sciences, Beijing
100049, China}

\begin{abstract}
Panchromatic spectral energy distribution (SED) from the ultraviolet (UV), optical to infrared (IR) photometry of
NGC 628, combined with the evolutionary stellar population synthesis, is used to derive the spatially resolved
age, metallicity and reddening maps. These parameter distributions show that the bulge of this galaxy is a
disk-like pseudobulge, which has the S{\'e}rsic index close to the exponential law, rich gas, and a young
circumnuclear ring structure. We also discover the disk has two distinct regions with different radial age and
metallicity gradients. The inner region is older and has a much steeper age gradient than the outer region of the disk.
Both these two regions and the central young structure can be seen in the radial profile of the optical color.
Based on the age and reddening distributions, we consider that the pseudobulge and disk are likely to have grown
via the secular evolution, which is the redistribution of mass and energy through the angular momentum transport
caused by the non-axisymmetric potential of the spirals. However, possible gas accretion events could affect
the outer region of the disk, due to abundant H{\sc i} gas accumulating in the outer disk.
\end{abstract}


\section{INTRODUCTION}
Galactic evolution at the early universe was dominated by the hierarchical clustering and merging that are
violent and rapid, secular evolution will be most important in the far future,
and now both processes might be equally fundamental \citep{ko04}. Recently, more and more observations have
revealed that the bulge with the S{\'e}rsic index approximating the exponential law might be grown via
the secular evolution of the disk \citep{ko93, bo02, ko04, fi08, ga09, fi09}. This kind of bulge is called
pseudobulge, which is very common in late-type galaxies \citep{bo02}. Pseudobulges have smaller S{\'e}rsic
indices than that of the \citet{de48} law in classical bulges, central gas concentration, active star formation,
nuclear rings, bars and/or spirals \citep{ko04}. Non-axisymmetric gravitational potential by the bar, oval, and/or
spiral structures rearranges the angular momentum and mass, makes gas infall, and triggers the central star
formation to create a pseudobulge \citep{re03, ko04}. The gas infall from the disk and the formation of the
bulge are secular. In addition to the secular evolution, distant encounters, gravitational interactions in clusters,
and gas accretions from the neighboring galaxies can affect the formation and evolution of the bulge and disk
\citep{to72,no88,co06,bo02b}. Thus, isolated nearby galaxies, which have large apparent sizes and are far from
the tidal interactions of other galaxies, are excellent objects to investigate the secular evolution of the
disk and formation of the bulge.

Spatially resolved age, abundance and reddening distributions can help us to understand the galactic evolution.
They provide some basic information about the star formation history, chemical compositions and evolution, and
interstellar medium. Both \citet{ko00} and \citet{li04} adopted different evolutionary population synthesis (EPS)
models to analyze the structure and evolution of a nearby Sab spiral galaxy (M81) using the photometric data of
multiple intermediate-band filters from 3000 {\AA} and 10000 {\AA} in the Beijing-Arizona-Taiwan-Connecticut (BATC)
multicolor sky survey. EPS has become a very popular technique to study the properties of stellar populations
in galaxies \citep{fo97,le99,br03,ko09} and it has been applied in many fields such as star clusters, galaxies, and
galactic clusters \citep{ab96, ko00, ma09a}.

In this paper, we will use the EPS model to derive the population properties (age, metallicity and intrinsic
reddening) and gain the clues for the formation and evolution of a nearby face-on SA(s)c galaxy, NGC 628. This galaxy,
also named M74, has an apparent size of about 10{\arcmin} and it is isolated \citep{ka92}. Light decomposition and
color distributions reveal that this galaxy has a disk-like bulge (i.e., pseudobulge) \citep{na92, co94, sa11, ga09}.
Besides the fifteen intermediate-band observations taken with the 60/90 cm Schmidt telescope of National
Astronomical Observatories of China as also used by \citet{ko00} and \citet{li04}, there are plenty of on-line
archival data for this galaxy from various surveys and telescopes, such as the Galaxy Evolution Explorer (GALEX),
Two Micron All Sky Survey (2MASS), and Spitzer space telescope. These panchromatic photometric data-set from
UV to IR can considerably aid us to obtain more accurate estimations of the properties (age, metallicity and
intrinsic reddening).

Previous studies on NGC 628 present various properties in respect of colors, stellar populations,
abundance, and reddening, etc., which can be compared with our results. The color of the whole galaxy is very blue,
ranging from the bluest of M33's colors to the bluest of M81's in UV - $V$ \citep{co94}. \citet{co94} reported that
the disk might have undergone significant star formations over the past 500 Myr by comparing the UV/optical colors
of stellar population synthesis models with those of NGC 628. \citet{na92} highlighted that two different stellar
populations dominate the inner and outer disk of the galaxy. \citet{sa11} presented the age and abundance profiles
calculated by the integral field spectroscopy in an unprecedented high spatial resolution. They found a young circumnuclear
star formation region near the galactic center. Distinct components with different ages and metallicities can be
easily discriminated in those profiles. \citet{ho76} and \citet{ke80} both used H$\alpha$ images to obtain
several hundreds of H{\sc ii} regions. Numerous H{\sc ii} regions, tending to lie along the spiral arms,
indicate that a number of stars have been forming recently in this nearby galaxy. Both \citet{ta83} and \citet{be92}
discovered a notable radial metallicity gradient according to the spectrophotometry and narrow-band imaging of H{\sc ii}
regions. Reddening in \citet{be92} does not display any trend as a function of the radial distance from the galactic center.
\citet{re06} investigated the 8 $\mu$m emission from the polycyclic aromatic hydrocarbons (PAHs) for some
disk galaxies (including NGC 628) and declared the 8 $\mu$m PAH surface brightness could be used as an approximate
tracer of the interstellar medium.

The paper is organized as follows. In Section \ref{data}, the observations from the BATC survey and the archival data
are described. Detailed data reduction is given in Section \ref{proc}. The introduction of the EPS model and fitting
method is provided in Section \ref{mode}. Decomposition of the surface brightness profile and the distributions of age,
metallicity and reddening as well as relevant comparisons are presented in Section \ref{para}. Discussions and conclusions
are made in Section \ref{diss} and \ref{conc}, respectively.

\section{OBSERVATIONS} \label{data}
NGC 628, as one of the nearby galaxies in the BATC sky survey \citep{bu94}, was observed by the
60/90 cm Schmidt telescope, which is deployed at the Xinglong Station belonging to National Astronomical
Observatories of China. A 2048$\times$2048 Ford Aerospace CCD with the pixel scale of 1.7{\arcsec} (15
$\mu$m pixel$^{-1}$) is mounted at the focal plane of the telescope (the focal ratio is f/3). The size of
field of view (FOV) is about 58{\arcmin} $\times$ 58{\arcmin}. For the photometric system, 15 intermediate-band
filters, covering the wavelength range of 3300 -- 10000 {\AA} with bandwidths of about 200 -- 300 {\AA}, are well
designed to avoid strong emission lines of the sky light \citep{fa96}. The observation mode is divided into two
parts: deep exposures for observations of the objects; short exposures at photometric nights for both the objects
and standard stars to do flux calibrations \citep{zh01}. The normal observations for all filters started in 1995
November and ended in 2004 February. Because of the bad image qualities of $a, b$, and $c$ filters, we reobserved
NGC 628 in 2009 winter for these three bands. Table \ref{tab1} summarizes the basic parameters of filters, the total
exposure time for each band, and some information of calibrations.

\begin{figure}
\epsscale{1.0}\plotone{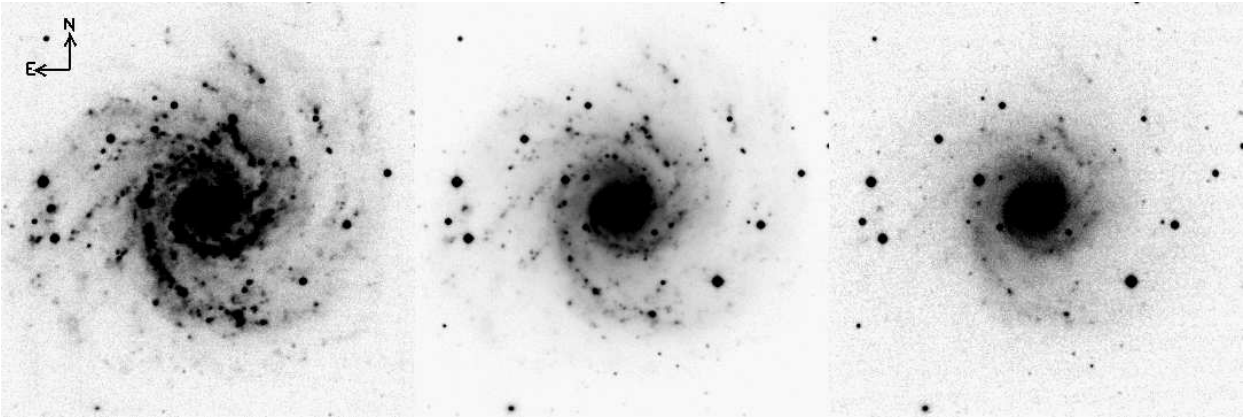} \caption{Images of NGC 628 in three
BATC bands centered at the wavelengths of 3890 (left), 6660 (middle), and 9190
{\AA} (right). The image scale is 361$\times$361 (1.71\arcsec/pixel). }
\label{fig1}
\end{figure}

Figure \ref{fig1} displays the combined images of NGC 628 in three BATC bands with wavelengths centered on 3890,
6660, and 9190 {\AA}. The galaxy is a grand-designed spiral galaxy with two symmetric inner arms and multiple long
and continuous outer arms (classified as Type 9 in the arm classes of \citet{el87}). We can see in those figures that
two clear spiral arms extend from the center to the disk. Some H{\sc ii} regions can be clearly recognized near the spiral
arms in the 6660 {\AA} band (this band covers the H$\alpha$ emission line) and in the 3890 {\AA} (near the ultraviolet
regions where massive young stellar populations generate strong emissions), while the stellar mass distribution of the
galaxy is visibly reflected in the near-infrared wavelength of 9190 {\AA}. The disk inclination is about 6$^\circ$
\citep{sh84,ka92} and the position angle of the major axis is about 25$^\circ$. The redshift is about 0.0022 \citep{hu99} and the
redshift-independent distance is about 8.6 Mpc \citep{he08}. This distance together with the disc inclination and
position angle is adopted throughout our paper.

In addition to the optical data of the BATC survey, the archival data of the ultraviolet and infrared observations are
collected to obtained the panchromatic spectral energy distribution of the galaxy from the on-line database. In the
Nearby Galaxy Survey (NGS) of GALEX, NGC 628 was imaged in the far-UV (FUV) and near-UV (NUV) bands, providing
substantially accurate ultraviolet observations with photometric precision up to 0.05 and 0.03 mag in the
FUV and NUV bands, respectively \citep{ma05, mo07}. Due to its bad fitting to the model of stellar population synthesis,
whose spectral libraries lack extreme horizontal-branch stars in the old stellar populations as mentioned in
\citet{ma09b}, the GALEX FUV band is excluded in the subsequent analysis. XMM Optical Monitor (XMM OM) also observed
this galaxy in three other ultraviolet bands (UVW1, UVM2, and UVW2) \citep{ku08}. The signal to noise ratio (SNR)
of the UVW2 band is extremely low, so this band is not used in our study. The UVM2 band together with UVW1
reinforces the observation data in the near-UV region. Near-infrared observations of NGC 628 in 2MASS $J$, $H$, and
$K_s$ bands \citep{sk06} are gained. Since only a small core in the $K_s$ band (radius less than half
an arcmin) has enough SNR and the public data of this band from the 2.3 m Bok telescope \citep{mc01} is available, we
replace the 2MASS $K_s$ observation by the $K_s$ image of the Bok telescope for a detailed description of the
data reduction refer to \citet{kn04}). Infrared observations in 3.6 and 4.5
$\mu$m of the Infrared Array Camera (IRAC, see \citet{fa04}) of the Spitzer space telescope \citep{we04} expand
the infrared data of NGC 628. Other bands in this telescope, such as 5.8, 8.0, 24, 70, and 160 $\mu$m, are located
in the radiation regions of gas and dust \citep{ke03}. Our study puts emphasis on the star light emitted
from the stellar content, so these bands are not used in the analysis of the stellar populations. The ultraviolet,
optical and infrared information of those telescopes and surveys including the instruments, images, and
corresponding references is listed in Table \ref{tab2}.

\section{Data Reduction} \label{proc}
\subsection{Processing Flow of BATC Images} \label{batc-pro}
Subtraction of the bias and dark current and the flat-field correction were performed
by the PIPLINE 1 procedure as described in \citet{fa96} once the CCD frames were observed.
Astrometric information was appended to the image header during this step. Deep frames
of each band were combined after shift and rotation to the same image center and
orientation according to the star positions. Cosmic ray hits and bad pixel effects were
fixed up during the image combination. Short exposures for both the object and the standard
stars were taken at photometric nights. We use these images to perform the flux calibration
for the above deep combined image as follows: 1) Four Oke-Gunn standard stars \citep{ok83}
are extracted in short exposure images of the standard stars and their instrumental magnitudes
are measured by a stellar photometric procedure (DAOPHOT; \citet{st87}); 2) Comparing the
instrumental magnitudes and calibrated AB magnitudes of these standard stars, we get atmospheric
extinction coefficient and instrument zeropoint; 3) Tens or hundreds of unsaturated bright stars
are selected as secondary standard stars in a short exposure image of the object, which is observed under the same
photometric condition as the standard stars; 4) according to the extinction coefficient and zeropoint,
the instrumental magnitudes of these stars are converted to flux-calibrated AB magnitudes. 5) The
average difference between the instrumental magnitudes of the same stars in the deep combined image and
their AB magnitudes are calculated; 6) with this difference, an analog-to-digital unit (ADU) can be converted
calibrated flux density unit of $\rm ergs ~s^{-1} cm^{-2} Hz^{-1}$. Calibration error of
each short exposure for the object contains two items: one is from the magnitude estimations of the
standard stars and the other is from the differences of those secondary standard stars between
instrumental magnitudes and AB ones. The total calibration error for each band as shown in the
last column of Table \ref{tab1} is the root mean square (RMS) of the calibration errors in all
short exposures, the number of which is presented in the seventh column of Table \ref{tab1}.

\subsection{Background Subtraction and Image Scaling}
Sky backgrounds should be removed from the mosaics of all bands in order to acquire
the intrinsic flux of the galaxy. For the GALEX mosaics, background images are provided,
but the background near the center is overestimated if we check their sky background
intensity images. The mosaics from the 2MASS, Bok $K_s$ band, and Spitzer were
constructed after removal of the sky background by their own techniques (see references
in Table \ref{tab2}). We obtain the smoothing sky background maps for all other mosaics by the
polynomial fitting method, based on the remaining background pixels after masking the source
signals extracted by SExtractor \citep{be96} and the galaxy by circling it with temperate
apertures for different bands. These fitted sky backgrounds are then removed from the original
mosaics.

Due to the diverse pixel scales ranging from 0.75{\arcsec} to 1.7 {\arcsec} and different
directions of the mosaics, we adjust them to the same scale and direction as the BATC images
(i.e., 1.7{\arcsec} per pixel, top of north and left of east). Take the Spitzer mosaic
for example, the pixel size is 0.75{\arcsec} but that of the BATC mosaic is 1.7{\arcsec}.
We first calculate the equatorial coordinate for each pixel of the BATC image, and
then find the corresponding pixel position in the Spitzer mosaic according to its astrometry.
After counting how many Spitzer pixels (can be decimals) are covered by the area of 1.7$\times$1.7
arcsec$^2$ centered at this pixel position, we sum all the fluxes of those pixels and
subsequently the adjustments of both the pixel scale and direction are completed simultaneously.
During the adjustments, we transform the pixel values of all bands to AB top-of-atmosphere
fluxes in unit of $\rm 10^{-30}ergs~s^{-1}cm^{-2}Hz^{-1}$. For the Bok 2.2 $\mu$m image from the
NASA/IPAC Extragalactic Database (NED), since its astrometry is not the same as ours and the magnitude
zeropoint is uncertain, we do something special: several stars with astrometric positions in the BATC
image are used to update the astrometry of this image; mosaic fluxes are recalibrated to the AB magnitude
system by some selected bright star in the same field from the 2MASS catalogues.

\begin{figure}
\epsscale{0.8}\plotone{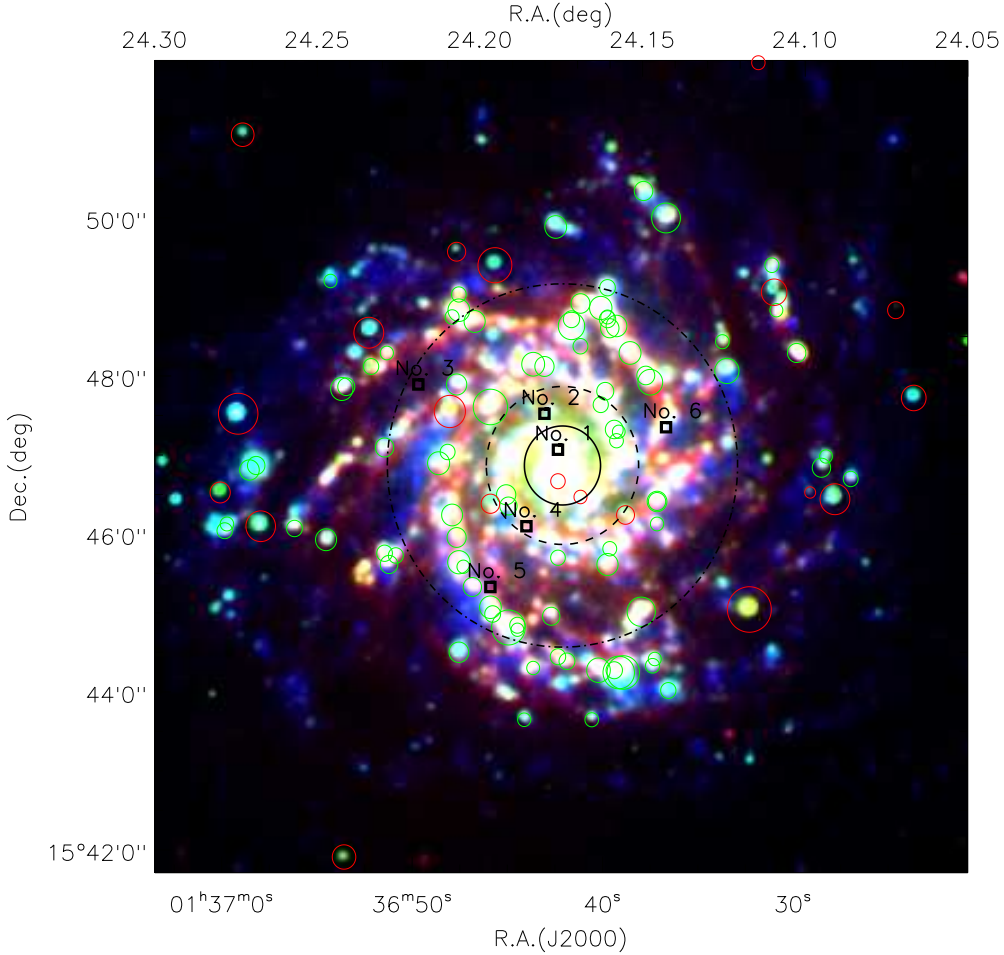} \caption{True color map of NGC 628 synthesized
with the 8 $\mu$m band (red), H$\alpha$ narrow band (6563 \AA, green) and NUV
band (2271 \AA, blue) images. These bands are chosen to present the young stellar
population and dust emission. Red and green solid circles are stars and
H{\sc ii} regions to be removed. Only those regions with radius larger than 5
arcsec in the catalogue of \citet{fa07} are displayed in this figure. Black squares
are the selected pixels for illustrating the model fitting goodness in section \ref{demo}.
The solid circle around the center, the dashed circle, and the outmost dash-dotted
circle represent the bulge together with the transition zone within $R < 0.5${\arcmin}, the
outer boundary of the inner region of the disk at 1 arcmin, and the outer boundary of
the outer disk at about 2.2{\arcmin} where the valid SEDs are still available (see Section
\ref{deco}), respectively.} \label{fig2}
\end{figure}

Figure \ref{fig2} displays the true color image of NGC 628, combined by three scaled mosaics
of the NUV, narrow H$\alpha$ (from NED), and 8 $\mu$m bands, which are chosen to present
the distributions of young stellar population and dust. From this image, we can find
that the whole disk is very blue, implying much younger populations in its disk than in the
bulge. Numerous H{\sc ii} regions, identified as luminous spots, mostly lie along the spiral
arms. Light in H{\sc ii} regions comes from both the old galactic background stellar
population and young stars, such as O and B type, which have been born recently as a simple
stellar population. These regions are more complex than the galaxy itself and they will
be specially treated in our series papers. So pixels belonging to 376 H{\sc ii} regions in
all mosaics are obliterated according to the catalogue of \citet{fa07}. We also find numerous
filaments and substructures generated by the PAH emission in red color of Figure \ref{fig2},
indicating the gas/dust distribution across the disk in addition to the strong PAH radiation
along the spiral arms.

\subsection{Image Smoothing and SED Extraction} \label{smooth}
Mosaics are smoothed for reasons that: 1) different astrometric precisions for
telescopes lead a single pixel to represent adjacent slightly different parts of the
galaxy; 2) transformations and rotations induce position errors; 3) different
observational conditions cause diverse image qualities (e.g., CCD noises and seeing);
4) SNR varies from the center to the edge of the galaxy so that the photometric
accuracy decreases in low SNR areas. Smoothing can improve SNR but in return reduce
the spatial resolution of the galaxy. We use the boxcar averaging method to smooth
all the mosaics as expressed in the following formula:
\begin{displaymath}
   M_{I,J} = \frac{1}{w^2}\sum_{K=I-\frac{w}{2}}^{I+\frac{w}{2}} \sum_{L=J-\frac{w}{2}}^{J+\frac{w}{2}} F_{K,L},
\end{displaymath}
where $I$ and $J$ specify the pixel position in the image, $M_{I,J}$ is the smoothed flux of
this pixel, $F_{K,L}$ is the original flux of any given pixel ($K$,$L$), and $w$ is the
smoothing window width (so called boxcar width), depending on the SNR of the specified
pixel in the BATC $j$ band image. The SNR ($R$) of a pixel is defined as
\begin{displaymath}
 R = \frac{S}{N} = \frac{S}{\sqrt{S+\delta^2}},
\end{displaymath}
where $S$ is the pixel flux, N is the noise, and $\delta$ is the standard deviation
of the global sky background. Here, SNR is similar to the nominal definition in a CCD
image, and we only use it to calculate the boxcar width and estimate relative photometric
errors for different bands. The window size for each pixel is determined by
$w = 2{\min}\{[\frac{R_m}{2R}], 5\} + 1$ where $[\frac{R_m}{2R}]$ is the minimum
integer larger than $\frac{R_m}{2R}$ and $R_m$ is set to 10. The formula promises $w$
to be an odd number to obviate the smoothing bias due to asymmetry. It means that
if SNR of the pixel is equal or larger than 10, the minimum smooth width is 3 pixels
and if SNR is equal or less than 1, the maximum width is 11.

After smoothing the mosaics, we extract the observed SED for each pixel. The observed SED
includes AB magnitudes of 23 bands from UV to IR, which are calculated as
\begin{equation}
m^{\rm obs} = -2.5{\log}_{10} S -48.6, \label{equ1}
\end{equation}
as long as the flux signal $S$ is larger than $3\delta$. The magnitude error ($\sigma$) is determined
by
\begin{equation}
 \sigma = 2.5{\log}_{10}(\frac{S+N}{S})=2.5{\log}_{10}(1+\frac{1}{R}). \label{equ2}
\end{equation}
At least 15 out of 23 bands with valid magnitudes are required for each SED to fit with model ones
as explained in the next section. As a result, we get 33242 SEDs and their errors for the whole
galaxy (see Table \ref{tab3} about the SEDs of 6 randomly sampled pixels).

\section{STELLAR POPULATION SYNTHESIS AND FITTING METHOD} \label{mode}
Evolutionary population synthesis has become a popular technique in studying different properties of
stellar populations and the evolution histories of galaxies, since there were some significant progresses
on the theory of stellar evolution \citep{re81,ma89,ki90,sc92,be94}, especially, the diverse phases of
evolution \citep{ib83,sc83,va93,gr93} and the completeness of stellar observational and theoretical spectral
libraries \citep{le03,pi98,le97}. Most fashionable synthesis models, like PEGASE \citep{fo97}, StarBurst99
\citep{le99}, GALEXEV \citep{br03}, and GALEV \citep{ko09}, are used to study globular clusters \citep{ma09a,ba00},
galaxies\citep{ko00,pa01,li04}, and galactic clusters \citep{ab96}, etc.

In the series papers on the study of nearby galaxies based on the photometric observations of the BATC
multicolor sky survey program, \citet{ko00} first quantified the chemical abundance, age and reddening
distributions of M81 using the synthesis models of SSP (from \citet{br93}). \citet{li04} then used those
spectral energy distributions (13 intermediate band filters) to gain the age and metallicity distributions
of M81 with the help of PEGASE, which contains exponentially decreasing star formation rates
(SFRs). Comparison of the above two models are described detailedly in their papers. We intend to apply the
PEGASE model to investigate the distributions of the stellar population properties in NGC 628.

\subsection{PEGASE Model}
PEGASE is a spectrophotometric evolution model for starburst and evolved
galaxies of the Hubble sequence attributed to the extension to the NIR atlas of
synthetic spectra and revised stellar libraries including cold stellar
parameters, stellar tracks of asymptotic giant branch (AGB), and post-AGB  phase
\citep{fo97}. The stellar tracks they adopt are mainly from the ``Padova" group,
which contain a wide range of chemical abundances, $Z$ = 0.0001, 0.0004, 0.004,
0.008, 0.02, 0.05, and 0.1 with $Y$ = 2.5$Z$ + 0.23 ($Z_\odot$ = 0.02, where $Z$
is the metallicity abundance and $Y$ is the helium abundance). The initial stellar
mass in the tracks ranges from 0.6$M_\odot$ to 120$M_\odot$, where in the track of
$Z$ = 0.1 pseudo-tracks with mass larger than 9$M_\odot$ are generated from the
corresponding masses in the $Z =0.02$ and $Z = 0.05$ tracks.

The stellar spectral libraries for optical wavelength (3130-10800\AA) come
from the stellar spectrophotometric atlas of \citet{gu83} which includes 175 stars
with complete ranges of spectral type and luminosity class. For the far-UV
(1230-3200\AA), stellar spectra are extracted from the International Ultraviolet Explorer
(IUE) ESA/NASA libraries \citep{he84}. In the extreme-UV (220-1230 \AA), spectra for
the effective temperate $T_{\rm eff} < 50000$ K are calculated using the models of
\citet{ku92}. The models of \citet{cl87} are used at all wavelength for hotter
stars ($T_{\rm eff} > 50000$ K). For cold stars dominating the
NIR (1 - 5 $\mu$m) of the spectra,  observational libraries when possible and
synthetic spectra otherwise are adopted. Blackbody radiations supplement the NIR wavelength
range of hotter stars. In the mid-infrared ($> 5 \mu$m) the analytic extension of
\citet{en92} is used for stars colder than 6000 K.

Simple stellar population (SSP) is regarded as a simple system whose stars are born at
the same time and with the same initial chemical compositions (typical example:
globular clusters). SSPs can be calculated in PEGASE with initial mass functions
(IMFs), stellar libraries, and stellar evolutionary tracks. PEGASE provides
several commonly applied IMFs, and the \citet{sa55} law with stellar mass larger
than 0.1 $M_\odot$ and less than 120 $M_\odot$ is used to generate SSP models in
this paper.

Galaxies are much more complex systems than globular clusters. Evolutionary composite
stellar population (CSP) can more exactly describe the evolution history of galaxies.
CSP is considered as a superimposition of SSPs of different ages. The integrated
spectrum of CSP is synthesized with an IMF and a SFR for every metallicity. We assume
each pixel of the mosaics to be such a CSP system that is composed of stars formed in
different periods. There are three kinds of SFRs in PEGASE and the exponentially
decreasing SFR with $\tau = 15$Gyr is used in our paper. Here, $\tau = 15$Gyr is
typical for Sc galaxies \citep{bo00}.

Another feature of PEGASE is that the model takes into account the nebular
emission (continuum and lines) generated by the ionized gas in the star-forming
regions. Nevertheless, the regimes used to synthesize the spectral energy
distribution for H{\sc ii} regions are very outdated. We do not consider the
nebular emission in the model spectra.

\subsection{Fitting Method}
For model spectra, both age from 0 to 20 Gyr and metallicity from 0.0001 to 0.1 are
interpolated by the PEGASE code in the logarithmic space with steps of 0.01. The \citet{sa55}
IMF and exponentially decreasing SFR are used to convolve these spectra to create CSPs of the
specified metallicity. We use the dust extinction model of \citet{ca89} to redden the CSP spectra
with a reddening step of 0.01 ranging from 0 to 1.0. By convolving the reddened CSP spectra with
the filter transmission curves of all bands, we can obtain the model SEDs set in the ranges of
different ages, abundances, and reddenings. The model SED contains convolved AB magnitudes of 23
bands computed as
\begin{equation}
  m_i^{\rm csp} = -2.5{\log}_{10}\frac{\int_\lambda{F_\lambda}(t, Z, E){T_i(\lambda)}d\lambda}
                  {\int_\lambda{T_i(\lambda)}d\lambda} - 48.6, \label{equ3}
\end{equation}
where $i$ is the specified band index, $T_i$ is the corresponding filter transmission curve,
$F(t, Z, E)$ is the CSP spectrum of the specified age $t$, metallicity $Z$ and reddening value
$E$ in $E(B-V)$, and $m_i^{\rm csp}$ is the resultant synthetical AB magnitude.
By comparing the observed SED (see Section \ref{smooth}) with the synthetical model
SEDs, we can simultaneously derive those three parameters ($t$, $Z$, and $E$) for each pixel
of NGC 628 via the $\chi^2$ minimum fitting method as described in \citet{ko00} and \citet{ma09b}:
\begin{displaymath}
 \chi^2 (t,Z,E) = \sum_{i=1}^{23}\frac{[m^{\rm obs}_i -
m_i^{\rm csp}(t, Z, E)]^2}{\sigma_{i}^2},
\end{displaymath}
where $m^{\rm obs}_i$ is the observed magnitude of the specified filter in the observed SED
as shown in Equation \ref{equ1}, $\sigma_{i}$ is the observed magnitude errors as given in
Equation \ref{equ2}, and $m^{\rm csp}_i(t, Z, E)$ is the corresponding synthetical
magnitude in the model SED as shown Equation \ref{equ3}.

The degenerate effect of age, metallicity, and dust reddening exists when we
match the observed SEDs with the model ones predicted by EPS, since the ways of
these three parameters to affecting spectra are similar. However, dust and gas
absorb the UV radiation of star light and reradiate out in the infrared region.
The multi-band photometric data from ultraviolet to infrared may aid in
degrading this degeneracy. Before being fitted by the above method, the observed
SEDs are corrected with the foreground reddening of the Galaxy by using the same
reddening law of \citet{ca89}. NGC 628 is located far away from the
Galactic disk as the Galactic latitude of its center is b = -45.7$^\circ$
(R.A.: 24.17$^\circ$ and Dec.: 15.78$^\circ$). So the foreground Galactic
reddening is small, about 0.07 mag in $E(B-V)$ \citep{sc98}.

\subsection{Fitting Results of Several Selected Pixels} \label{demo}
We randomly sample 6 pixels from different parts of NGC 628 to illustrate the SEDs, fitted
parameters, and the fitting goodness: No. 1 near the galactic core, No. 2 from the inner region
of the disk, No. 3 from the outer region of the disk, No. 4 from the inner spiral arm, No. 5
from the outer spiral arm, and No. 6 close to an H{\sc ii} region (see Table \ref{tab3}). The
positions of these six pixels are marked in the true color map of Figure \ref{fig2} and the observed
SEDs, matched model SEDs, and the corresponding model spectra are plotted in Figure \ref{fig4}.
Excellent match to the model spectrum can be seen in this figure. We might coarsely conclude that:
the core is oldest; age becomes younger distant from the center; spiral arms are younger than the
rest components except for the places near the H{\sc ii} regions with much younger ages and richer
abundances.

\section{Structure and Stellar Population Analysis of NGC 628} \label{para}
Based on the spectral synthesis model and fitting method as described in the previous section,
we obtain spatially resolved age, metallicity and intrinsic reddening distributions of NGC 628 after
fitting the observed SED of each pixel with the models. In order to analyze the properties of different
components, we first decompose the galaxy into two components in both the optical and near-infrared
surface photometry. Then the stellar population analysis of the whole galaxy and its components is
presented in the following sections.

\subsection{Bulge-disk Decomposition of the Surface Brightness Profiles}
As generally known in spiral galaxies, the bulge and disk exhibit many physical and
dynamical differences. The bulge is usually brighter, older, and dynamically hotter than
the disk. In order to analyze different components of NGC 628 with respect to the two-dimensional
features, we disassemble the galaxy into two main components. These two components
are obtained by fitting the azimuthally averaged surface brightness with the exponential
disk profile \citep{fr70} and the \citet{se68} law following the iterative decomposition
procedure of \citet{ko77}. In this decomposition method, two separate regions of the disk and
bulge are chosen in the beginning. Then, a least square fitting of an exponential law
to the radial surface brightness in the disk range is performed. The calculated exponential
disk contribution is extrapolated to the bulge range and subtracted from the observed
profile to get a first estimate of the bulge component, which has been fitted by the \citet{se68}
law. In the same way, this fitted \citet{se68} law is extrapolated to the disk range and
subtracted from the observed profile to get an estimate of the underlying disk, which be fitted by
the exponential law again. The procedure repeats until all the parameters of two laws
converge within a specified accuracy ($10^{-3}$).

The radial profile is the corrected azimuthally-averaged values,
given the disk inclination of 6$^\circ$, the position angle of the major axis of
25$^\circ$, and the distance of 8.6 Mpc as previously mentioned. The value at each radius
is defined as the average of the pixel values (here, surface brightness) within a suitably
specified annulus of that radius centered on the galaxy nucleus. The profile error is
calculated as the standard deviation of those pixel values. All radial profiles of different
parameters (age, color, metallicity, and reddening) in the rest of the paper are computed in
the same way with those adopted parameters (inclination, position angle and distance).

\begin{figure}
\epsscale{1.0}\plotone{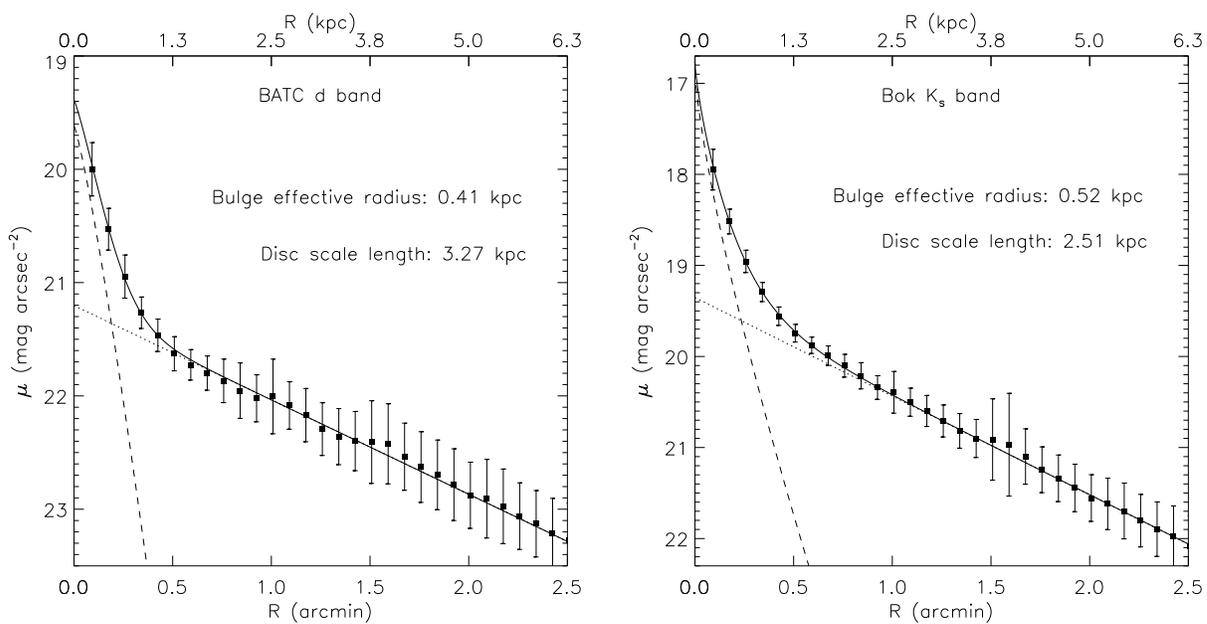} \caption{ Left: radial surface brightness profile of the BATC $d$ band
as plotted in filled squares with error bars. Right: radial surface brightness profile of the BATC $K_s$ band
as plotted in filled squares with error bars. The dashed curve and dotted line in both panels are the fitted
\citet{se68} law for the bulge and exponential law for the disk, respectively. The solid curve is the sum of
these two fitted components. \label{fig3}}
\end{figure}

\begin{figure}
\epsscale{1.0}\plotone{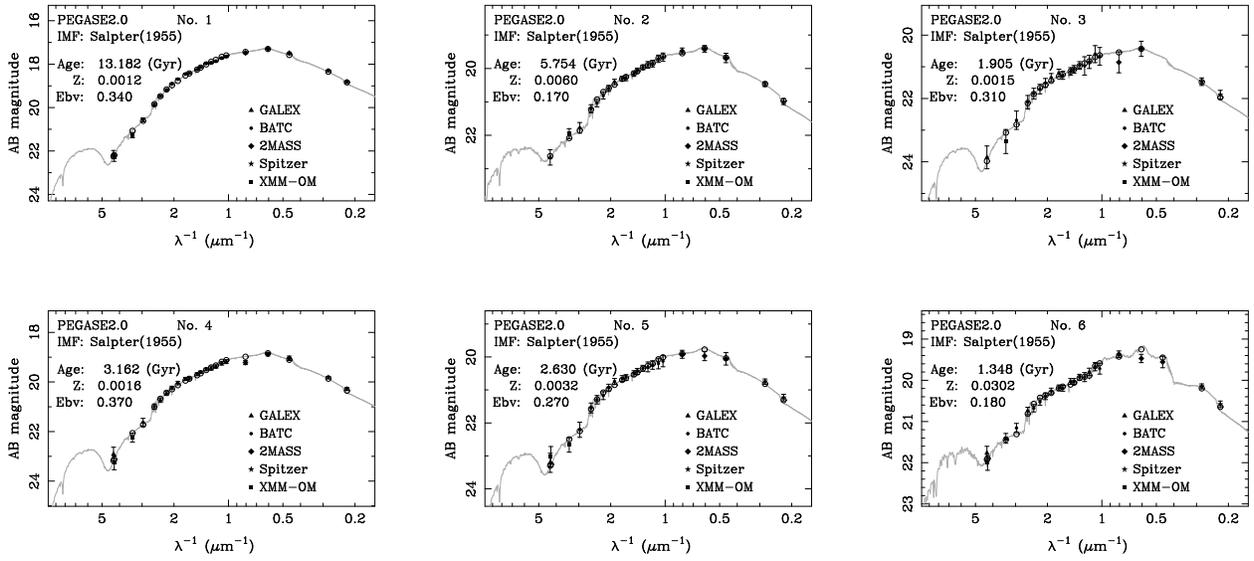}
\caption{Combined plots of observed SEDs, matched model SEDs and corresponding
spectra for six randomly sampled pixels as described in Table \ref{tab3}. The
filled symbols are the observed AB magnitudes of specified bands and open
circles are the matched model magnitudes. Spectra are drawn in grey. Vertical
bars are the measured errors of the observed magnitudes as defined in
Equation \ref{equ2}.\label{fig4}}
\end{figure}

Left panel of Figure \ref{fig3} shows the radial surface brightness profile of the BATC $d$
band (filled squares with error bars), whose effective wavelength is close to the broad $B$
band, and the decomposed bulge and disk components as shown in dashed curve and dotted line.
We use the same regions where the bulge and disk clearly dominate the observed profile as
adopted in the paper of \citet{bo81} to derive the decompositions iteratively: between 3.5{\arcsec} and
23.0{\arcsec} for the bulge and between 69.0{\arcsec} and 230.2{\arcsec} for the disk.
The decomposition procedure gives the central surface brightness ($\mu_0^B$) of 19.64 $\pm$
0.19 mag arcsec$^{-2}$, the effective radius ($r_e$) of 0.41$\pm$ 0.04 kpc (0.16 $\pm$ 0.02
arcmin), the S{\'e}rsic index ($n_b$) of 0.82 $\pm$ 0.08 for the S{\'e}rsic law of the bulge, the disc
central surface brightness $\mu_0^D = 21.19 \pm 0.11$ mag arcsec$^{-2}$ and scale length
$h = 3.25 \pm 0.20$ kpc (1.30 $\pm$ 0.08 arcmin) for the exponential law of the disk.
We obtained a much smaller effective radius of the bulge than that of \citet{bo81},
because the \citet{de48} law can not fit the bulge component very well when we check Figure
6 of their paper. Although large discrepancy for the bulge, the scale length of the disk
approximate those of \citet{bo81} corresponding to their adopted distance of 12.2 Mpc. Actually,
many exceptions when fitting the bulge with the \citet{de48} law were mentioned by \citet{mc09}
(and references therein), and it is more acceptable that the projected three-dimensional bulge
profile is fitted by the generalized \citet{se68} law.

Although blue bands are used to derive the structural parameters of different components
traditionally, their light profiles are substantially affected by the recent star formation
and the extinction of the gas and dust. To the contrary, the NIR luminosity is a good tracer
of the stellar mass and maps the distribution of the old stellar populations in galaxies. It
is insensitive to the luminosity of young massive stars and unaffected by the extinction of
the gas and dust for its ignorable extinction coefficient. Therefore,
we also decompose the surface brightness profile of the NIR $K_s$ band into two components
as shown in the right panel of Figure \ref{fig3}. The central surface brightness, the effective
radius, and the S{\'e}rsic index of the bulge are 16.87 $\pm$ 0.18 mag arcsec$^{-2}$, 0.52 $\pm$
0.04 kpc (0.21 $\pm$ 0.02 arcmin), and 1.31 $\pm$ 0.11, respectively. The disc central surface
brightness and scale length are 19.35 $\pm$ 0.10 mag arcsec$^{-2}$ and 2.51 $\pm$ 0.12 kpc (1.00 $\pm$ 0.05 arcmin),
respectively. The effective radius, S{\'e}rsic index, and disc scale length are consistent with the structural
parameters derived by \citet{ga09} for the HST $H$ band photometry \citep{ga09}, which give
$r_e = 0.21 \pm 0.002$ arcmin, $n_b = 1.23 \pm 0.01$, and $h = 1.18 \pm 0.02$ arcmin. From the fitting parameters, we find
that the disk profile decrease more dramatically in the NIR band than in blue optical band,
since there are a number of young stellar populations located in the disk and they raise relatively
more luminosity in the blue bands. The effective radius of the $K_s$ band is larger than that
of the $d$ band, giving a more reliable span of the stellar mass of the bulge. We regard the
region of $R < 0.21${\arcmin} as the region of the bulge, the region of $R > 0.5${\arcmin}
where the profile starts to deviate from the exponential law as the region of the disk, and
the region between these two areas as a transition zone.

\subsection{Age Distribution} \label{agedist}
\begin{figure}
\epsscale{1.0}\plotone{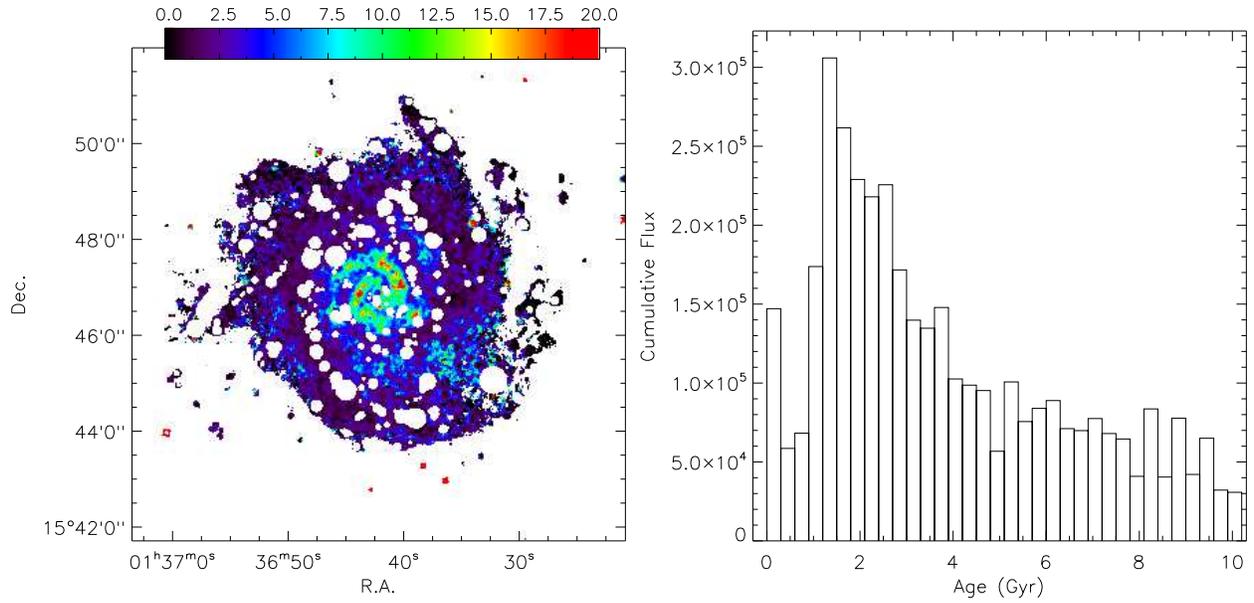} \caption{ Left: two-dimensional age distribution of NGC 628. Blank regions are
masked stars and H{\sc ii} regions. Right: histogram distribution of mass weighted ages. The cumulative flux in the
vertical ordinate is the total apparent luminosity of the $K_s$ bands in $\rm 10^{-30} ergs ~s^{-1} cm^{-2} Hz^{-1} arcsec^{-2}$
which represents the stellar mass.  \label{fig5}}
\end{figure}

Figure \ref{fig5} displays the age distributions of NGC 628 in 2D (left panel) and histogram (right
panel). In the left panel of Figure \ref{fig5}, we find that age is oldest near the galactic
core and becomes younger with the increasing radial distance (the age ranges from about 10
Gyr to 2.0 Gyr). Two apparent spiral arms in the disk, where considerable H{\sc ii} regions are concentrated,
are much younger ($\sim$ 1.3 Gyr) than any other components. They extend from the center to the outer
region of this galaxy. The inter-arm areas are filled with relatively older stellar populations.

In order to derive the statistical properties of the stellar populations in more physical meaning,
the near-infrared luminosity is used as the mass weight for each pixel in the galaxy image. Since
the near-infrared luminosity is insensitive to young massive stars, and reddening in this wavelength
range can be also ignored, the near-infrared bands, especially the $K_s$ band, are ideal tracers of the
stellar mass \citep{co01, ko01}. In the following of this paper, we will derive various statistical
features (e.g., average values, histograms, and radial profiles) of age, metallicity, and reddening
weighted by the $K_s$ band flux.

The average age of the whole galaxy, bulge and disk is about 4.9 Gyr, 7.5 Gyr and 4.4 Gyr, respectively.
Half of the total stellar mass is younger than about 3.5 Gyr as seen in the right panel of Figure \ref{fig5}.
In this figure, we can also see that there are quite a few extremely young stellar populations with age
less than 1 Gyr. A majority of them are located close to the double spiral arms and some H{\sc ii} regions
as confirmed by their positions on the age map.

\citet{sa11} presented a wide-field IFS survey on NGC 628 and obtained 2D spectra in the FOV of about
6 arcmin in diameter with spatial resolution of about 2.7 arcsec. A dithering mode was adopted for the
central pointing to increase the spatial resolution. They derived the global age of about 8.2 Gyr by
both fitting the integrated spectrum of the whole observed galaxy to linearly combined SSPs and
averaging the radial age distribution. The discrepancy of the average age with respect to our estimates
mainly originates from the different size of the observed galactic area, which in our study is more than 2
times that of \citet{sa11}. So considerable rather young stellar populations in the outer region of the disk are
involved in the calculation of our average age.

\subsection{Two Distinct Disc Components} \label{deco}
\begin{figure}
\epsscale{1.0}\plotone{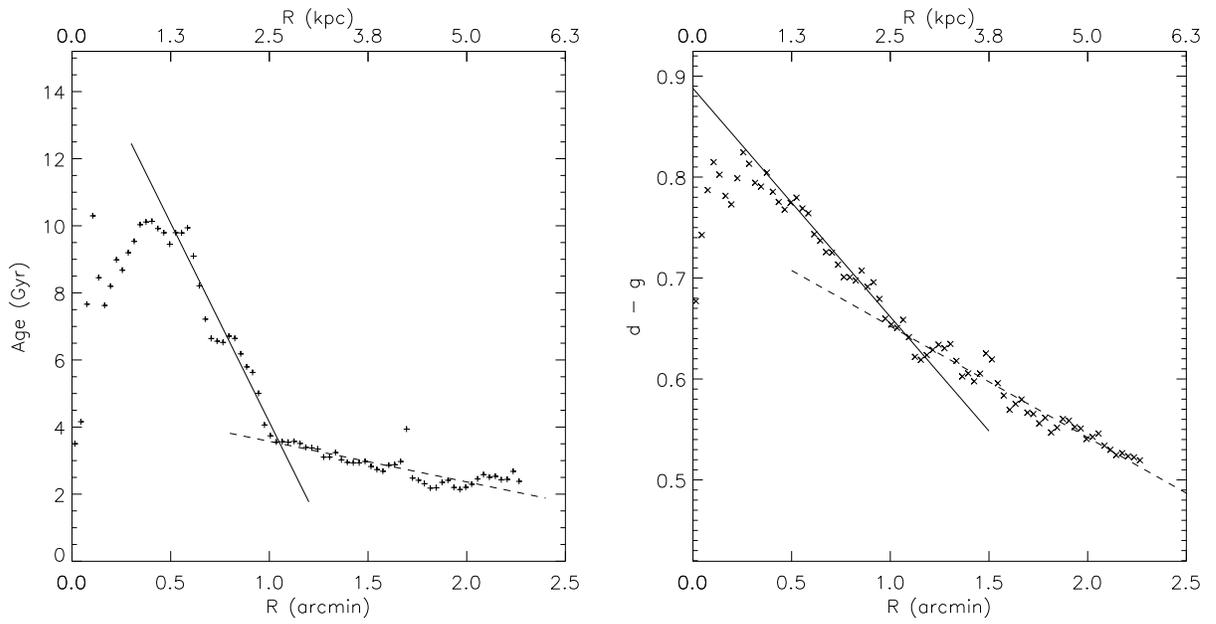} \caption{ Left: radial age profile. Right: radial profile of the BATC $d - g$ color.
The solid and dashed lines present two fitted gradients in the inner region of the disk (0.5 -- 1.0{\arcmin}) and
the outer region of the disk (1.0 -- 2.2 {\arcmin}), respectively. \label{fig6}}
\end{figure}

The radial age profile in the left panel of Figure \ref{fig6} shows that there are two distinct components
in the disk, where we consider the radius range of 0.5 -- 1.0{\arcmin} as the inner region of the disk and the
range from 1.0{\arcmin} to 2.2{\arcmin} as the outer region of the disk (separated at the radius of about 1.0{\arcmin}).
This radial age profile is very similar to that of \citet{sa11} within the radius they analyzed regardless
of the system uncertainties. The inner region of the disk is older and its age gradient is much steeper than that
the outer region. The mean ages of these two components are about 7.2 Gyr and 2.9 Gyr, respectively, and the
slopes are $11.9 \pm 1.0$ Gyr arcmin$^{-1}$ (4.8 Gyr kpc$^{-1}$) for the inner region in the solid line and
$1.2 \pm 0.1$ Gyr arcmin$^{-1}$ (0.5 Gyr kpc$^{-1}$) for the outer region in the dashed line of the left panel
of Figure \ref{fig6}.

Such distinct disc components in the age distribution should appear in the color profile which is
related to the properties of the underlying stellar population, such as age and metallicity.
We calculated the radial profile of the BATC $d - g$ color as plotted in the right panel of Figure \ref{fig6}.
These two bands are selected according to their effective wavelengths approximating the broad bands of
$B$ and $V$. The variation of this color profile is similar to that of the age profile: a flattening or even
inverse tendency within 0.5{\arcmin} and two different gradients in the region of $R > 0.5${\arcmin}.

Although the inner and outer regions are both in the exponential disk, the age of the inner region
is much older than that of the outer one. The steeper age gradient of the inner region of the disk
indicates this smaller component should experience a very long evolution history in an age span of about 7 Gyr.
The outer region of the disk with much shallower gradient might be formed 2 -- 3 Gyr ago within a time interval
of about 1 Gyr, which implies that NGC 628 might endure some encounter or accretion events and large number
of gas fell in to produce new stars and generate the large-scale outer region of the disk in a very short time
(see Section \ref{disk}).

\subsection{Young Circumnuclear Ring Structure}
In the left panel of Figure \ref{fig6}, contrary to the disc components, an inverse age gradient
resides in the bulge and the transition zone (also discovered in the $d - g$ color profile),
while the oldest age up to 10 Gyr comes up at the galactocentric radius of about 0.5{\arcmin} (1.3 kpc)
instead of the galactic center. Relatively younger structure near the galactic
core indicates that some processes like gas inflow directly through the galactic disk to the bulge
might occur in the history and trigger the star formation. This may be also associated to
the possible existence of a weak barred potential which induces the circumnuclear star formation as presented in
the paper of \citet{se02}.

\citet{na92} showed the broad band $UBVRI$ surface photometry of NGC 628, and found an obvious blue drop
in their $B-V$ profile for $R < 1.5$ kpc, which can be predicted by models due to infall of gas in the
bulge. Sub-mm CO (1-0) observation of \citet{wa95} and infrared 2.3 $\mu$m CO
absorption spectroscopy of \citet{ja99} displayed the existence of a circumnuclear star forming
ring in the center of the galaxy. \citet{co94} used the UV surface photometry and discovered that
the nuclear region has the overall morphological characteristics of spiral arm materials resembling
those of M33 and might endure a significant star formation in the past few Gyrs predicted by the UV-optical
colors.

In the paper of \citet{sa11}, in addition to the decreasing age gradient in the inner region and
possible shallow gradient in the outer region of the disk, an inverse age gradient was also obtained in a
circumnuclear ring at about 25{\arcsec}, which is very close to the position of our result.
This kind of ring structure was also detected by \citet{ga06} in their H$\beta$ and
[O {\sc iii}] distributions, obtained by the integral field spectroscopy with the SAURON
Integral Field Unit pointing to the galactic core (about 33\arcsec$\times$41\arcsec).

\subsection{Metallicity Distribution}
\begin{figure}
\epsscale{1.0}\plotone{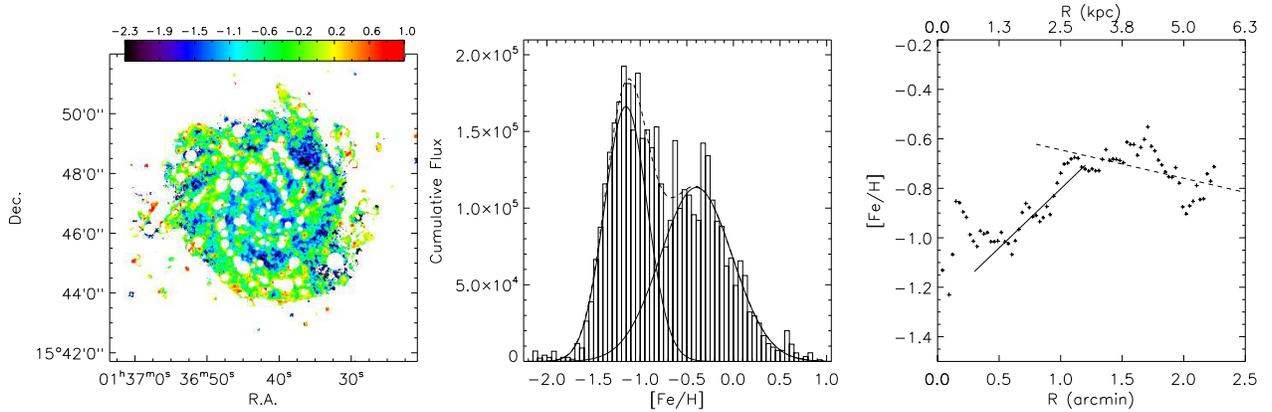} \caption{Left: two-dimensional [Fe/H] distribution of NGC 628.
Blank regions are masked stars and H{\sc ii} regions. Middle: histogram distribution of mass weighted
metallicity. The cumulative flux in the vertical ordinate is the total apparent luminosity of the $K_s$ bands in
$\rm 10^{-30} ergs ~s^{-1} cm^{-2} Hz^{-1} arcsec^{-2}$. Two components of poor and rich metallicity
are fitted by a double Gaussian function (bimodal), which is plotted in the dashed curve. The solid
curves are two single Gaussian items of the bimodal. Right: radial [Fe/H] distribution. The solid line and
dashed lines show the [Fe/H] gradients within the range of 0.5 -- 1.0{\arcmin} and the range of
1.0 -- 2.5{\arcmin}, respectively.  \label{fig7}}
\end{figure}

The metallicity $Z$, derived by SED fit, is converted to [Fe/H] using the standard chemical composition of the Sun \citep{gr98}.
Related histogram and radial distribution are also calculated. In the left panel of Figure
\ref{fig7}, we can see that the metallicity is lower near the core than in the outer regions and the
abundance close to spiral arms is much richer than the inter-arms due to their poles of the
frequently born place of massive stars.

Average [Fe/H]s of the whole galaxy, bulge, disk, and the inner and outer region of the disk
are about -0.77, -1.04, -0.78, -0.93, and -0.71 dex, respectively. The derived global metallicity
is lower than that of \citet{sa11}, whose spectra mostly lie along the spiral arms and in the bulge.
Much richer abundances of H{\sc ii} regions in the spirals make the global metallicity of \citet{sa11}
larger than ours. In the histogram of [Fe/H] (middle panel of Figure \ref{fig7}), two components
dominate this spiral galaxy: the older stellar population with poorer abundance and younger population
with richer metallicity. The poor stellar population mostly lies within the galactic core and inter-arms,
while the rich one mainly resides close to the spiral arms and the outer region of the disk.
The distributions of these two populations take on a bimodal profile. We fit the overall distribution
by a double Gaussian function which has two single Gaussian functions as its items. Parameters of these
two Gaussian components are summarized in Table \ref{tab4}. The two components and the bimodal curve are
respectively plotted in solid and dashed lines in the middle panel of Figure \ref{fig7}.

As presented by H{\sc ii} observations of \citet{ta83} and \citet{be92}, an obvious radial O/H abundance
gradient exists in the disk of NGC 628, which is 0.21 dex arcmin$^{-1}$ and $0.17 \pm 0.004$ dex arcmin$^{-1}$
, respectively corresponding to their adopted distances. Equally, a weak decreasing [Fe/H] gradient located
in the outer region of the disk is shown in the right panel of Figure \ref{fig7}. \citet{sa11} also obtained
the radial stellar metallicity distribution, the profile of which is very similar to that of ours.
The slope of this weak gradient is about $0.11 \pm 0.03$  dex arcmin$^{-1}$ (0.05 dex kpc$^{-1}$). Comparing
those gradients with our result of $0.11$ dex arcmin$^{-1}$, we discover that the stellar metallicity gradient
is somewhat lower than that of the gas abundance. It is notable that there is a reverse gradient of the inner
region of the disk with a slope of $0.48 \pm 0.08$ dex arcmin$^{-1}$ (0.19 dex kpc$^{-1}$). Such kind of
abundance distribution in the disk provides fundamental constraints on the chemical evolution of NGC 628.

\subsection{Intrinsic Reddening Distribution}
\begin{figure}
\epsscale{1.0}\plotone{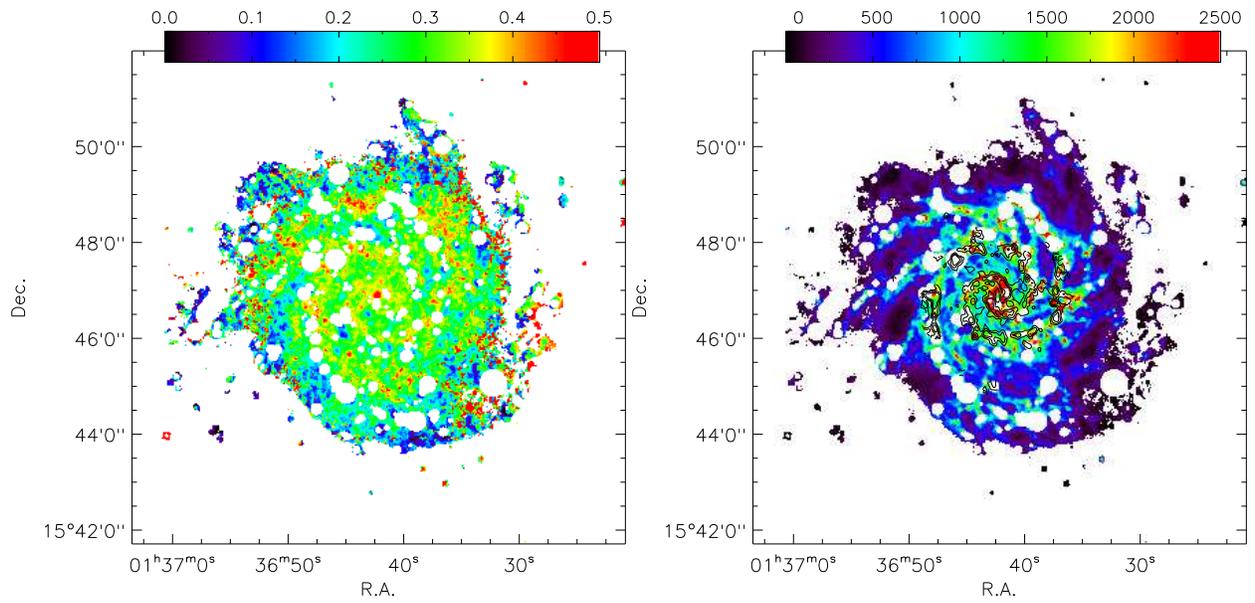} \caption{Left: two-dimensional intrinsic reddening
distribution of NGC 628. Blank regions are masked stars and H{\sc ii} regions. Right:
IRAC 8.0 $\mu$m stellar flux-subtracted image in the same area as the reddening map.
Overlapped contours show CO intensities (contour levels: 1.5, 3, and 5 $\rm Jy beam^{-1} km s^{-1}$). \label{fig8}}
\end{figure}

Both dust and gas absorb and scatter star light and make the galaxy redder. Extinctions
are much larger near the UV bands than in the IR bands. Accurate photometry from
the UV to infrared can derive reliable intrinsic reddening ($E(B-V)$) of NGC 628
as shown in the left panel of Figure \ref{fig8}. In this map, regions close to the
spiral arms and the bulge hold relatively larger reddenings. The mean reddening of
those regions close to spiral arms is about 0.38 and show no radial gradient, which
is consistent with the observational measurements in \citet{be92}. Some inter-arm
regions and parts of the outmost disk appears to be less reddened by dust and gas.

In order to confirm the reliability of the reddening value, dust emission of the
Spitzer 8.0 $\mu$m is compared with the reddening distribution. We construct the
intensity map from the dust in Spitzer 8.0$\mu$m (right panel of Figure \ref{fig8}) dominated by the
PAH emissions using the method of \citet{he04}, who subtracted the background stellar
fluxes from the observed fluxes. In the determination of the stellar fluxes in longer
wavelengths than 3.6 $\mu$m, \citet{he04} extrapolated the 3.6 $\mu$m flux using the
stellar population model of StarBurst99 \citep{le99} to obtain the scale factor of
0.232 in 8.0 $\mu$m (i.e., $D_{8.0} = I_{8.0} - 0.232I_{3.6}$, where
$D_{8.0}$ is the intensity of the dust emission and $I_{3.6}$  and $I_{8.0}$ are the observed
intensities in the Spitzer 3.6 and 8.0 $\mu$m, respectively). Superposed contours in the right
map of Figure \ref{fig8} show the distribution of 2.6 mm CO (1-0) emission intensity from the
Berkeley-Illinois-Maryland Association Survey of Nearby Galaxies (BIMA SONG) \citep{re01}.
\citet{re06} concluded that the CO and 8 $\mu$m emission morphologies are similar as
also presented in this map and 8 $\mu$m PAH surface brightness can be used as a
possible tracer of the interstellar medium. Clearly strong PAH emissions preferentially distribute
along the spiral arms where a number of H{\sc ii} regions radiate UV photons to photodissociate
the PAHs around the molecule clouds and generate IR emissions. Compared with the reddening map,
we discover that many of the regions with large reddening values in the galaxy show luminance
peaks in the 8 $\mu$m PAH image, implying the existence of abundant gas and dust. Some other
regions with large reddenings might be caused by the absorption of dust lanes as checked in the optical
observations with respect to the positions of primary dust lanes and reddening feathers, which
were investigated by \citet{la06}.

\begin{figure}
\epsscale{1.0} \plotone{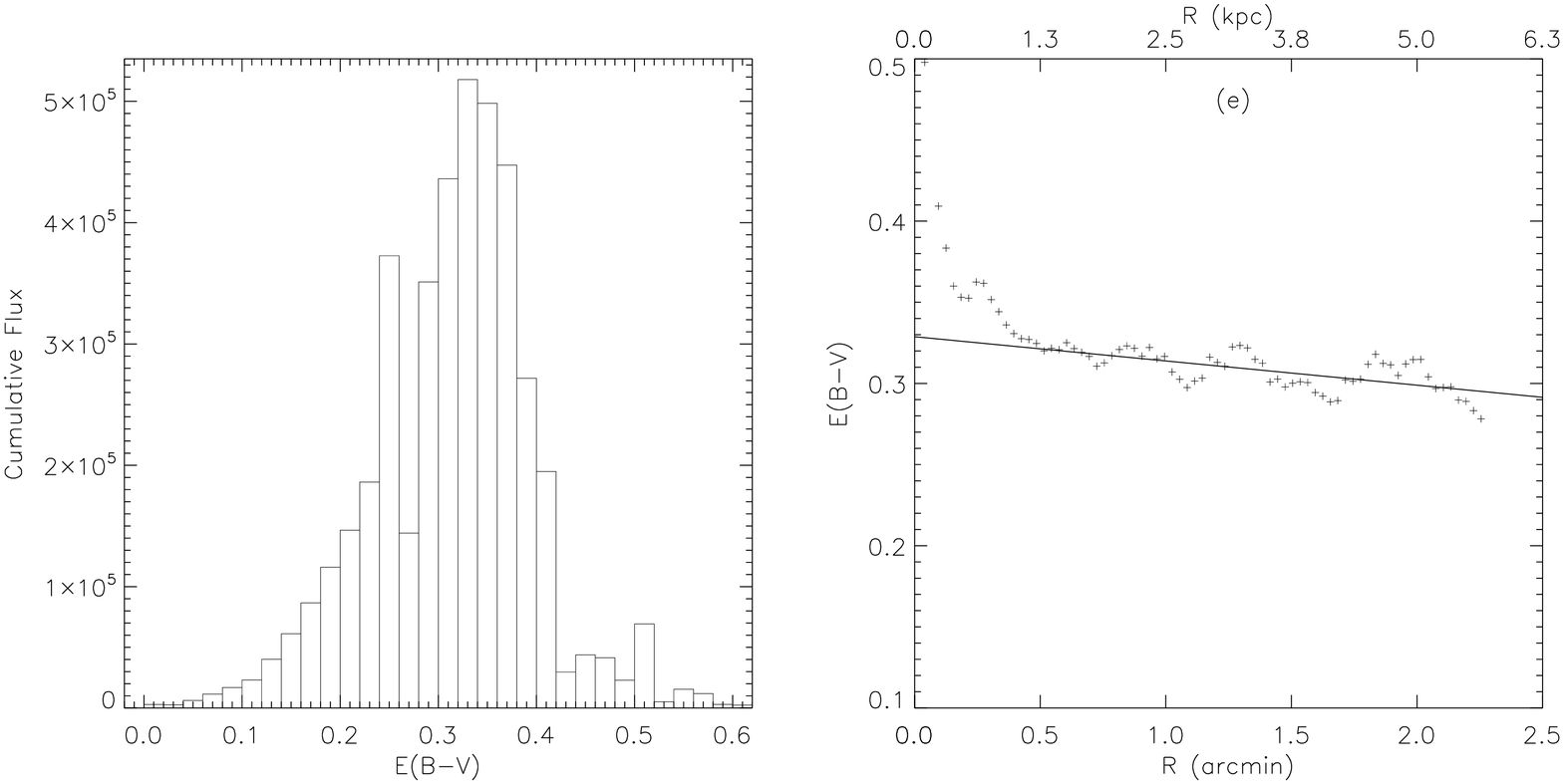} \caption{Left: histogram distribution of the overall
intrinsic reddening in $E(B-V)$. The cumulative flux in the vertical ordinate is the total apparent
luminosity of the $K_s$ bands in $\rm 10^{-30} ergs ~s^{-1} cm^{-2} Hz^{-1} arcsec^{-2}$.
Right: radial distribution of the intrinsic reddening. The solid line presents the fitted radial gradient. \label{fig9}}
\end{figure}

Histogram in the left panel of Figure \ref{fig9} shows that $E(B-V)$ varies from 0.1 to 0.6 and
the mean reddening is about 0.31. Average reddenings of the bulge, the disk, and the inner and outer
regions of the disk are about 0.41, 0.31, 0.32, and 0.31 mag, respectively. The radial distribution in the right
panel of Figure \ref{fig9} presents a very weak gradient of the intrinsic reddening with a slope
of 0.015 $\pm$ 0.002 mag arcmin$^{-1}$. Several bumps in this plot are produced by the large reddening
values near the spiral arms. We can see in the radial profile that the bulge as well as the transition
zone has larger reddening values than the outside area, supporting the young circumnuclear ring
structure holding abundant gas and dust. That is, a considerable volume of dust and gas should be
deposited in the core of NGC 628, giving a rise of star formation in its recent history. In contrast,
\citet{ko00} studied another nearby galaxy (M81) of Sab type using a similar method. Their results
show that M81 has a much lower reddening range of 0.08 -- 0.15 mag in the bulge than ours and the
disk reddening ($\sim$ 0.2 mag) is also somewhat smaller, implying that NGC 628 might be a gas-rich galaxy.

\section{Discussions} \label{diss}
\subsection{Pseudobulge and Secular Evolution}
The S{\'e}rsic indices in both optical and NIR bands as we have derived from the brightness profile fitting are
close to 1, which represents the exponential law. It has been increasingly discovered in recent years that
bulges with the surface brightness profiles close to the exponential law are significantly different from
those with larger S{\'e}rsic indices resembling the elliptical galaxies \citep{ko93, ko04, fi08, ga09, fi09}.
Now, two types of bulges are typically called: classical bulges that are dynamically hot and featureless and
pseudobulges that retain the memory of the disk origin. The pseudobulges are very common in late-type galaxies
(75\% of 77 Sd--Sm galaxies as given in \citet{bo02}). They have flatter shapes, kinematics dominated by rotation
and corresponding smaller velocity dispersion, active star formation, nuclear bars, nuclear rings, and/or
nuclear spirals and nearly exponential brightness profiles ($n_b \sim 1$), while the classical bulges are
dominated by random motions, contain old stellar populations, and are more similar to the E-type galaxies
\citep[][and references therein]{fi08, ga09, fi09}. Specifically, \citet{fi09} identified pseudobulges as
those bulges containing nuclear bars, nuclear spirals, and/or nuclear rings and having the S{\'e}rsic index
less than 2.

The S{\'e}rsic indices of NGC 628 are 0.83 for the blue band and 1.31 for the NIR band, both of which approximate
the exponential law. Nuclear spiral arms are clearly seen in the UV band of this galaxy \citep{co94}. Moreover,
the young ring structure detected in our age map and cental concentration of rich gas shown in the reddening map
declare recent young star formation in the galactic center. Therefore, the bulge of NGC 628 belongs to that so-called
disk-like pseudobulge.

Unlike the classical bulges, which are typically merger-built, pseudobulges have the opportunity to be grown via
the internal secular evolution of the disk \citep{ko04,fi09}. Disk galaxies can evolve by the rearrangement
of mass and angular momentum driven by the non-axisymmetries of bars, ovals and/or spiral structures, which actuate the
gas infall towards the galactic center, trigger the star formation, and build up the central mass concentrations of
the pseudobulges \citep{ko04}. In barred or oval disk galaxies, gravitational torques and/or shocks near their potential
minimum cause the gas inflow and outflow, which can bring together the disk gas, hence trigger the star formation and
make outer rings, inner rings, and central mass concentrations (pseudobulges) \citep{re03,fi09}. In normal spiral galaxies
without bars and ovals, spiral structures are maintained by the density wave that propagate through the disk. Shocks are
formed as gas approaches and leaves the arms, making the gas lose the energy and sink into the center, and form the disk-like
bulge \citep{ko04}. NGC 628 is a late-type spiral galaxy without evident bars and ovals, so it is possible that its pseudobulge
was formed by the secular evolution of the disk driven by the non-axisymmetric potential of spiral arms. Actually,
\citet{fi09} concluded that the median pseudobulges could have grown the current stellar mass at their present-day SFRs
in 8 Gyr and their results are consistent with a scenario in which bulge growth via internal secular star formation.
The age of the pseudobulge in NGC 628 as we obtained previously is about 7.5 Gyr, similar to the results
of \citet{fi09} and hence supporting the secular evolution.

Besides secular evolution, possible effects that build the pseudobulges include extremely gas-rich accretion events,
distant gravitational encounters, and gravitational interactions in a cluster \citep{ko04,fi08}. NGC 628 is an isolated
galaxies and no large galaxies are found within close distance which could generate possible tidal interactions
\citep{ka92}. Thus, gravitational encounters and interactions should be unimportant for the formation of the pseudobulge
in this galaxy. H{\sc i} observations by \citet{ka92} show two symmetrical high velocity complexes (HVCs) and an extended tail
to the southwest. Gas accretion events from an external H{\sc i}-rich object may form the outer tail, and accretion
can also create extreme velocity deviations making those HVCs visible. \citet{ka92} argued that we might be witnessing
the flattening of the disk after accreting nearby companions long time ago according to the orbit of the tail,
disc warp sustained by a steady gas infall, and complicated velocity distortion in the outer disk. However,
it is extraordinarily uncertain whether this accretion-induced formation of pseudobulges can retain the disk-like
properties. They also did not exclude the primordial origin that regards the observed gas distribution as the left-over
of the initial formation of the galaxy. Nevertheless, the primordial origin and external accretion of gas disk in NGC 628
do not prevent this galaxy from forming the pseudobulge by the processes in the secular evolution and the secular evolution
can still have an effect.

\subsection{Nucleus}
Both the age map (Figure \ref{fig5}) and the reddening map (Figure \ref{fig8}) show the nucleus with young age (about
3.5 Gyr) and rich gas (about 0.5 mag in $E(B-V)$). Decomposition of the NIR $H$ band in extremely high
resolution of the HST image also presents that the radial brightness profile exceeds the S{\'e}rsic law of the
pseudobulge in the innermost of the galaxy \citep{ga09}. This kind of compact nuclei hosts nuclear star
clusters and they are very common in late-type galaxies \citep{bo02}. The nuclei are obviously distinguished
from the ambient (pseudo)bulges and disk in the sense of a much smaller effective radius and higher effective
surface brightness. Stellar populations of nuclei tend to have blue colors implying young age for the stars
which contribute most of the light (typical example of M33 in \citet{lo02}). The nucleus and the pseudobulge
can both exist in the same galaxy but they show a great number of differences, making the nucleus harder to
be understood. Maybe these tiny and dense nuclei were formed via dynamical friction driving the clusters sinking
to the center as mentioned in \citet{ko04}. However, the origin and evolution of nuclei are quite complicated and its discussion
is far beyond the ability with current results of our study.

\subsection{Scenarios of Disk Evolution} \label{disk}
In our results, we discover that the disk of NGC 628 can be divided into two parts: the inner region and
the outer region. These two parts have different properties of stellar populations as far as age and chemical
abundance be concerned. Disk dynamics points that the inner part of the disk shrinks and the outer part expands
due to angular momentum transport caused by differential rotation and non-axisymmetric self-gravitating mode,
such as spiral arms \citep{ly79, li87, tr89}. This redistribution of mass and angular momentum results from the
minimization of the total energy with the total angular momentum conservation. The gas contraction in the inner
part of the disk could more easily trigger the star formation than in the outer part, and hence evolve more quickly. Merely
by this kind of internal secular evolution, the disk can construct two different regions with a large age discrepancy.
Certainly, one should confirm whether this process is fast enough to form the structures within the Hubble time.
Non-axisymmetry of the potential caused by the spiral arms can provide the engine for rapid evolution as discussed
in \citet{ko04}. On the other hand, H{\sc i} observations of \citet{sh84} and \citet{ka92} display that most of the gas concentrates
on the disk and an extensive tail lies to the southwest. Gas accretion events from an H{\sc i}-rich object with $M_{\rm HI}
\sim 9\times10^8 M_\odot$ might occur long time ago and they affect only the outer disk loosely bounded
by gravitation \citep{ka92}. Therefore, the accreted gas could rule the following star formation activities in the
outer region, forming a low stellar-density and young outer region of the disk.

\section{CONCLUSION} \label{conc}
Recently, evolutionary stellar population synthesis models become more and more popular
in studying the properties of the stellar populations in order to understand the star formation
and evolution histories of galaxies. In this paper, we adopt one of the EPS models,
PEGASE, to analyze the age, metallicity and reddening distributions of a nearby face-on
spiral galaxy (NGC 628) based on multi-band photometric data ranging from UV to IR.
On-line archival data from GALEX, XMM OM, 2MASS, Bok telescope and Spitzer provide the ultraviolet
and near-infrared photometry of this galaxy. The optical observations are obtained in the BATC
survey with 15 intermediate bands ranging from 3000 {\AA} to about 1 $\mu$m. By comparing
the photometric SED (totally 23 bands) with those of model ones calculated from
the PEGASE synthetical spectra, we derive the properties of stellar population for each part
of the galaxy (area of 1.7{\arcsec} in the BATC pixel size) and hence the spatially resolved age,
metallicity and reddening maps.

Structure parameters are calculated in optical and NIR bands by fitting the surface brightness
profiles with the \citet{se68} law for the bulge and exponential law for the disk, in order to
discuss the properties of different components in NGC 628. The bulge size is 0.52 kpc and disc
scale length is 2.51 kpc. The mean stellar age of the galaxy is about 4.9 Gyr, so young that
a great number of stars might form in the recent history of 2 -- 3 Gyr. The age of
the bulge is about 7.5 Gyr, presenting the oldest stellar population. The radial profile shows
the age becomes younger and younger from the center to the outer disk. We confirm the presence of a young
circumnuclear ring structure in the bulge and the transition zone within the galactocentric distance of about 0.5{\arcmin}.
This young structure behaves to have flat and even inverse age gradient, showing active star formation.
The disk can be divided into two parts as presented in the radial age profile: the old inner region and
young outer region, which have distinct age gradients of 11.9 Gyr arcmin$^{-1}$ and 1.2 Gyr arcmin$^{-1}$,
respectively. The mean ages of these two regions are about 7.2 Gyr and 2.9 Gyr. Both the circumnuclear
ring structure and two different regions of the disk can be clearly identified in the BATC
$d - g$ color profile.

Abundance map of NGC 628 in [Fe/H] reveals that the bulge is consisted of old stellar populations with low
metallicity ([Fe/H] $\sim$ -1.04 dex) and the disk is relatively richer ([Fe/H] $\sim$ -0.78 dex). A very
weak abundance gradient with slope of 0.11 dex arcmin$^{-1}$ (0.05 dex kpc$^{-1}$) for the outer region of
the disk is obtained in the radial profile, which is smaller than the gas abundance gradient obtained by
the measurements of H{\sc ii} regions. Much larger [Fe/H] values in the metallicity map are discovered
neighboring to H{\sc ii} regions, where numerous young massive stars have been being formed. Reddening map shows obvious
spiral-like lanes extending from the galactic core to the outer disk. Comparing with the IRAC 8 $\mu$m dust
emission image, we find that many parts of the reddening distribution correlate the PAH emissions. The global
average reddening in $E(B-V)$ is about 0.31. Reddening values of the bulge and disk are about 0.37 and
0.31, respectively. A tiny gradient of the radial reddening distribution is found, whose slope is 0.015 mag arcmin$^{-1}$.

At the end of this paper, the formation and evolution of different components of NGC 628 are discussed based on
the results. As very common in late-type galaxies, the bulge of this galaxy is a pseudobulge, which has the S{\'e}rsic index
close to the exponential law, young star formations and central spiral structures as seen in the UV band. The spiral
structures generate the non-axisymmetric potential, make the gas fall into the center and form the central
gas concentration which triggers the star formation and grows the pseudobulge. All these processes are
secular and our results about the pseudobulge support its growth by the secular evolution.
However, H{\sc i} gas distribution in the disk does not rule out the primordial origin and gas accretions
affecting the formation of the pseudobulge. We also discover that a nucleus hosting star clusters can be
obviously noticed, which has very young stellar populations and large reddening. Since NGC 628 is an isolated
galaxy, gravitational interactions with other external galaxies are excluded in both the bulge and disk
evolution. The secular evolution of the disk can form two distinct regions with a great age discrepancy,
although gas accretion if existing, could also affect the outer region.

\acknowledgments
We thank the referee for his/her thoughtful comments and insightful suggestions
that improve our paper greatly.
This work is based in part on observations made with the Spitzer Space
Telescope, which is operated by the Jet Propulsion Laboratory, California
Institute of Technology under a contract with NASA. This publication makes use
of data products from the Two Micron All Sky Survey, which is a joint project of
the University of Massachusetts and the Infrared Processing and Analysis
Center/California Institute of Technology, funded by the National Aeronautics
and Space Administration and the National Science Foundation. We thank the
SINGS (Spitzer), GALEX, and XMM-OM teams for making this research possible. We
thank the observations of the $K_s$ band by the Bok telescope and CO emission by
the BIMA SONG. This research has made use of the NASA/IPAC Extragalactic
Database (NED) which is operated by the Jet Propulsion Laboratory, California
Institute of Technology, under contract with the National Aeronautics and Space
Administration. This work was supported by the Chinese National Natural Science
Foundation grands No. 10873016, 10633020, 10603006, 10803007, 10903011, 11003021, and 11073032, and
by National Basic Research Program of China (973 Program), No. 2007CB815403.

\clearpage
\begin{table}
\scriptsize
\begin{center}
\caption{Filter information in the BATC photometric system and statistics of
observations}
\label{tab1}
\begin{tabular}{cccccccc}
\tableline\tableline No. & Filter & $\lambda_{\rm eff}$\tablenotemark{a}
(\AA) & Bandwidth(\AA) &
Exp\tablenotemark{b} (s) & FWHM\tablenotemark{c} ({\arcsec}) &
N\tablenotemark{d} & rms\tablenotemark{e} \\
\tableline
01 & $a$ & 3360 & 222 & 13200 & 4.61 & 2 & 0.041\\
02 & $b$ & 3890 & 187 &  8400 & 4.45 & 3 & 0.016\\
03 & $c$ & 4210 & 185 &  7200 & 4.01 & 6 & 0.008\\
04 & $d$ & 4550 & 222 & 11400 & 3.43 & 5 & 0.015\\
05 & $e$ & 4920 & 225 & 13800 & 3.73 & 3 & 0.009\\
06 & $f$ & 5270 & 211 & 11100 & 4.04 & 3 & 0.006\\
07 & $g$ & 5795 & 176 &  6000 & 4.44 & 2 & 0.002\\
08 & $h$ & 6075 & 190 &  6000 & 3.48 & 4 & 0.004\\
09 & $i$ & 6660 & 312 &  3900 & 4.28 & 5 & 0.005\\
10 & $j$ & 7050 & 121 &  6600 & 4.24 & 7 & 0.005\\
11 & $k$ & 7490 & 125 &  7500 & 5.07 & 5 & 0.009\\
12 & $m$ & 8020 & 179 &  7800 & 5.25 & 1 & 0.021\\
13 & $n$ & 8480 & 152 & 10800 & 4.44 & 5 & 0.005\\
14 & $o$ & 9190 & 194 & 15300 & 4.57 & 8 & 0.009\\
15 & $p$ & 9745 & 188 & 19900 & 4.89 & 1 & 0.020\\
\tableline
\end{tabular}
\tablenotetext{a}{Effective wavelength of each filter}
\tablenotetext{b}{Total exposure time of the combined image}
\tablenotetext{b}{Full width at half maximum (FWHM) of the combined image}
\tablenotetext{d}{Image number for calibration}
\tablenotetext{e}{Calibration error in magnitude as explained in Section \ref{batc-pro}}
\end{center}
\end{table}

\clearpage

\clearpage
\begin{table}
\begin{center}
\scriptsize
\caption{Multi-band observations from different telescopes for NGC 628}
\label{tab2}
\begin{tabular}{c|c|c|c|c|c|c|c|c}
\tableline\tableline
\multirow{2}{*}{Name\tablenotemark{a}} & \multirow{2}{*}{Filter
\tablenotemark{b}} & $\lambda_{\rm eff}$\tablenotemark{c} & Bandwidth
 & \multirow{2}{*}{FOV\tablenotemark{d}}  & scale\tablenotemark{e} &
Exp\tablenotemark{f} & FWHM\tablenotemark{g} &
\multirow{2}{*}{Reference}\\
 & & (\AA) & (\AA) & & ({\arcsec}) & (s) & ({\arcsec}) & \\
\tableline
\multirow{2}{*}{GALEX} & FUV & 1516 & 268 & 1.28$^\circ$ & 1.5 & 1636.05 & 4.3
& \multirow{2}{*}{1}\\
      & NUV & 2267 & 732  & 1.24$^\circ$ & 1.5 & 1636.05 & 5.3 & \\ \hline
\multirow{3}{*}{XMM-OM} & UVW1 & 2905 & 620 & 17{\arcmin} & 0.95   & $\sim$1000.0  &
2.0 & \multirow{3}{*}{2} \\
       & UVM2 & 2298 & 439 & 17{\arcmin} & 0.95   & $\sim$1000.0  & 1.8 & \\
       & UVW2 & 2070 & 500 & 17{\arcmin} & 0.95   & $\sim$1000.0  & 2.0 & \\ \hline
BATC  & $a$ -- $p$ & 3000 -- 9900 & 120 -- 310 & 58{\arcmin} & 1.7 & 1.1 -- 5h & $\sim$4.3 & 3\\
\hline
\multirow{3}{*}{2MASS} & $J$  & 1.235 $\mu$m & 0.162 $\mu$m & 0.39$^\circ$ & 1 & 7.8 &
$\sim$2.5 & \multirow{3}{*}{4}\\
      & $H$  & 1.662 $\mu$m & 0.251 $\mu$m & 0.39$^\circ$ & 1 & 7.8 & $\sim$2.5 & \\
      & $K_s$& 2.159 $\mu$m & 0.262 $\mu$m & 0.39$^\circ$ & 1 & 7.8 & $\sim$2.5 & \\
\hline
Bok   & $K_s$& 2.16 $\mu$m & 0.262 $\mu$m & 8.5{\arcmin} & 0.24 &
$\sim$1500 & 1.5 & 5 \\ \hline
\multirow{4}{*}{Spitzer} & IRAC1 & 3.550 $\mu$m & 0.75 $\mu$m &  5.2{\arcmin} & 0.75 &
1500.8 & 1.66 & \multirow{4}{*}{6}\\
        & IRAC2 & 4.493 $\mu$m & 1.01 $\mu$m &  5.2{\arcmin} & 0.75 & 1500.8 & 1.72 & \\
        & IRAC3 & 5.731 $\mu$m & 1.42 $\mu$m &  5.2{\arcmin} & 0.75 & 1500.8 & 1.88 & \\
        & IRAC4 & 7.872 $\mu$m & 2.93 $\mu$m &  5.2{\arcmin} & 0.75 & 1500.8 & 1.98 & \\
\tableline
\end{tabular}
\tablenotetext{a}{Name of the telescope or survey}
\tablenotetext{b}{Name of the filter}
\tablenotetext{c}{Effective wavelength in angstrom if not specified}
\tablenotetext{d}{Field of view of the CCD camera}
\tablenotetext{e}{Mosaic pixel scale}
\tablenotetext{f}{Total exposure time in unit of seconds if not specified}
\tablenotetext{g}{Full width at half maximum of the mosaic}
\tablecomments{References. (1) \citet{mo07}; (2) \citet{ku08}; (3) \citet{zh01};
(4) \citet{sk06}; (5) \citet{kn04}; (6) \citet{fa04}}
\end{center}
\end{table}

\clearpage

\clearpage

\clearpage

\thispagestyle{empty}
\begin{sidewaystable}
\tiny
\setlength{\tabcolsep}{0.3mm}
\begin{center}
\caption{Observed SEDs of several sampled pixels and corresponding fitted
parameters}
\label{tab3}
\makebox[\textwidth]{
\begin{tabular}{ccccccccccccccccccccccccccccccccccccccccccccc}
\tableline\tableline
No & X & Y & $\alpha$(J2000) & $\delta$(J2000) & NUV & UVM2 & UVW1 & $a$ &
$b$ & $c$ & $d$ & $e$ & $f$ & $g$ & $h$ & $i$ & $j$ & $k$ & $m$ & $n$ & $o$ &
$p$ & $J$ & $H$ & $K_s$ & IRAC1 & IRAC2 & Age & E(B-V) & Z &
[Fe/H] \\
\tableline
1&182&189&24.1739&15.7868&22.76&22.89&21.67&20.92&20.19&19.78&19.46&19.15&
18.91&18.65&18.61&18.42&18.30&18.12&18.05&17.98&17.82&17.73&17.54&17.33&17.52&18
.32&18.83&13.2&0.34& 0.0012&-1.18\\
\nodata&\nodata&\nodata&\nodata&\nodata&0.18&0.18&0.12&0.09&0.06&0.05&0.05&
0.04&0.03&0.03&0.03&0.03&0.03&0.02&0.03&0.03&0.03&0.03&0.02&0.02&0.03&0.03&0.03&
0.6&0.01&0.0001&0.04\\
2&188&205&24.1766&15.7945&23.48&23.90&22.65&21.91&21.32&20.98&20.66&20.38&
20.13&19.94&19.90&19.72&19.62&19.48&19.40&19.36&19.20&19.14&19.09&18.68&18.95&19
.73&20.24&8.5&0.30&0.0016&-1.09\\
\nodata&\nodata&\nodata&\nodata&\nodata&0.27&0.37&0.19&0.15&0.11&0.09&0.08&
0.07&0.06&0.06&0.06&0.05&0.05&0.05&0.07&0.06&0.07&0.07&0.06&0.06&0.08&0.05&0.06&
0.9&0.01&0.0003&0.08\\
3&244&218&24.2041&15.8014&24.84&99.99&23.65&23.06&22.41&22.28&21.96&21.79&
21.52&21.34&21.37&21.25&21.16&21.04&20.89&20.73&20.61&20.70&20.28&20.31&99.99&21
.27&21.66&1.1&0.43&0.0010&-1.23\\
\nodata&\nodata&\nodata&\nodata&\nodata&0.44&9.99&0.30&0.33&0.20&0.21&0.17&
0.14&0.12&0.11&0.12&0.12&0.12&0.12&0.28&0.16&0.18&0.29&0.17&0.26&9.99&0.10&0.12&
0.3&0.03&0.0005&0.20\\
4&196&155&24.1813&15.7708&23.52&23.86&22.65&21.98&21.32&21.04&20.74&20.49&
20.27&20.08&20.05&19.86&19.77&19.64&19.57&19.49&19.35&19.27&19.28&18.92&19.07&19
.83&20.29&3.2&0.37&0.0016&-1.04\\
\nodata&\nodata&\nodata&\nodata&\nodata&0.28&0.28&0.18&0.16&0.11&0.09&0.08&
0.07&0.07&0.06&0.06&0.06&0.05&0.05&0.08&0.07&0.07&0.08&0.08&0.07&0.09&0.05&0.06&
0.4&0.01&0.0003&0.09\\
5&212&128&24.1895&15.7582&23.82&23.62&23.07&22.56&21.87&21.60&21.45&21.18&
20.97&20.86&20.82&20.67&20.60&20.49&20.43&20.36&20.30&20.22&19.98&20.01&20.08&20
.76&21.24&2.6&0.27&0.0032&-0.74\\
\nodata&\nodata&\nodata&\nodata&\nodata&0.29&0.31&0.22&0.23&0.15&0.13&0.12&
0.10&0.09&0.09&0.09&0.08&0.09&0.09&0.15&0.13&0.16&0.17&0.12&0.14&0.19&0.08&0.10&
0.5&0.02&0.0010&0.19\\
6&134&199&24.1501&15.7909&22.37&22.60&21.81&21.51&21.07&20.98&20.80&20.64&
20.49&20.39&20.39&20.20&20.19&20.07&20.05&19.91&19.77&19.81&19.43&19.51&19.58&20
.16&20.59&1.3&0.18&0.0302&0.30\\
\nodata&\nodata&\nodata&\nodata&\nodata&0.16&0.18&0.12&0.11&0.09&0.09&0.08&
0.08&0.07&0.07&0.07&0.07&0.07&0.07&0.11&0.09&0.10&0.13&0.08&0.10&0.14&0.06&0.08&
0.2&0.02&0.0040&0.06\\
\tableline
\end{tabular}
}
\end{center}
{X and Y are the pixel position in the image with the size of $361\times361$. The positions
of these pixels are marked in Figure \ref{fig1}. Both
right ascension $\alpha$ and declination $\delta$ are in degrees. The last four
columns are the fitted parameters. There are two rows for each pixel: one row
contains the observed SEDs and fitted results and the other includes their
errors. Note that magnitudes of 99.99 and errors of 9.99 are set for the
bands whose SNRs are three times lower than the sky background standard
deviation.}


\end{sidewaystable}

\clearpage

\clearpage

\clearpage

\clearpage

\clearpage
\begin{table}
\begin{center}
\caption{Bimodal parameters of the [Fe/H] distribution fitted by a double Gaussian function}
\label{tab4}
\begin{tabular}{c|c|c|c|c|c|c}
\tableline\tableline
Abundance & $\mu$\tablenotemark{a} & Error$_\mu$\tablenotemark{b}  & $\sigma$\tablenotemark{c} & Error$_\sigma$&
$A$\tablenotemark{d}& Error$_A$ \\ \hline
poor & -1.15 & 0.03 & 0.23 & 0.03 & 1.66$\times10^5$ & 0.20$\times10^5$\\  \hline
rich & -0.40 & 0.08 & 0.39 & 0.07 & 1.13$\times10^5$ & 0.09$\times10^5$\\
\tableline
\end{tabular}
\tablenotetext{a}{Mathematical expectation of Gaussian function}
\tablenotetext{b}{Error (for all parameters) with 95\% confidence}
\tablenotetext{c}{Standard deviation of Gaussian function}
\tablenotetext{d}{Constant coefficient of Gaussian function}
\end{center}
\end{table}

\clearpage

\clearpage

\end{document}